\documentclass{article}
\usepackage{grffile}
\usepackage{arxiv}

\usepackage[utf8]{inputenc}
\usepackage[T1]{fontenc}

\usepackage{amsmath, amssymb, amsfonts, stmaryrd}

\usepackage{graphicx}
\usepackage{subcaption}
\usepackage[table]{xcolor}
\usepackage{booktabs}

\usepackage{algorithm}
\usepackage{algpseudocode}

\usepackage{url}
\usepackage{hyperref}
\usepackage[numbers,sort&compress]{natbib}
\usepackage{doi}

\usepackage{microtype}
\usepackage{nicefrac}

\title{Assessing the Influence of Locational Suitability on the Spatial Distribution of Household Wealth in Bernalillo County, NM}

\author{
Onyedikachi J. Okeke \\
Department of Geography and Environmental Studies\\
Texas State University, San Marcos,Texas, USA \\
\texttt{okaykaygeoinfo@gmail.com} 
\And
Uloma E. Nelson  \\
Department of Computer Science and Science Education\\
New Mexico Highland University, Las Vegas, NM, USA \\
\texttt{}
\And
Chukwudi Nwaogu \\
Department of Environmental Management\\
Federal University of Technology, Owerri, Nigeria \\
\texttt{}
\And
Olumide O. Oladoyin \\
Department of Chemical \& Biological Engineering\\
University of New Mexico, Albuquerque, NM, USA \\
\texttt{}
\And
Emmanuel Kubuafor \\
Department of Mathematics and Statistics\\
University of New Mexico, Albuquerque, USA \\
\texttt{}
\And
Dennis Baidoo \\
Department of Mathematics and Statistics\\
University of New Mexico, Albuquerque, USA \\
\texttt{}
\And
Titilope Akinyemi \\
MBA Business Analysis\\
Georgia State University, Atlanta, USA \\
\texttt{}
\And
Adedoyin S. Ajeyomi \\
Spatial and Data Science Society of Nigeria, Abuja, Nigeria\\
\texttt{}
\And
Rekiya A. Idris \\
Spatial and Data Science Society of Nigeria, Abuja, Nigeria \\
\texttt{}
\And
Isaac A. Fabunmi \\
Department of Teacher Preparation\\
New Mexico State University, Las Cruces, NM, USA \\
\texttt{}
}

\renewcommand{\shorttitle}{Assessing the influence of locational suitability}

\hypersetup{
  pdftitle={Assessing the influence of locational suitability},
  pdfauthor={Okeke, Nwaogu, Oladoyin, Kubuafor, Baidoo, Akinyemi, Ajeyomi, Idris, Ibukun, Fabunmi},
  pdfkeywords={Spatial inequality},
}

\begin{document}

\maketitle

\begin{abstract}
This study employs Multiscale Geographically Weighted Regression (MGWR) to analyze the spatial determinants of household wealth in Bernalillo County, NM. The model integrates socio-demographic, environmental, and proximity-based variables to assess how locational suitability influences economic outcomes. Key factors include income, home value, elevation, PM2.5 levels, and distance to essential services like schools, markets, and hospitals. The MGWR model performs robustly, explaining approximately 63\% of the variance in household wealth. Results reveal that proximity to markets, schools, and parks significantly enhances wealth in over 40\% of neighbourhoods. Conversely, closeness to hospitals and bus stops correlates negatively with wealth, suggesting associated disamenities can detract from housing appeal. Strong spatial autocorrelation (Moran’s I = 0.53, p < 0.001) confirms that wealthier households are significantly clustered, indicating economic conditions are influenced by localized factors. This research demonstrates that the relationships between locational suitability and wealth are spatially variable.
\end{abstract}


\section{Introduction}
Household wealth distribution reflects persistent spatial inequalities that shape social and economic outcomes across the United States. In Bernalillo County, New Mexico, patterns of affluence and deprivation align closely with locational suitability, including proximity to employment centers, transportation networks, and environmental quality. The county, anchored by Albuquerque, displays striking disparities between prosperous neighborhoods in the Northeast Heights and economically disadvantaged areas in the South Valley and International District \cite{Logan2018}. These geographic contrasts reveal how place influences access to opportunity, housing markets, and overall well-being. Understanding such spatial disparities is essential for designing equitable urban policies and addressing the geography of opportunity in mid-sized metropolitan regions \cite{Tsingos2020}.

Urban and regional studies have long recognized the link between housing, location, and economic well-being. Housing quality and affordability influence not only household stability but also intergenerational mobility and wealth accumulation \cite{Herbert2005}. Access to affordable homes in well-serviced neighborhoods enhances the potential for equity building and resilience to economic shocks \cite{Argiolas2014}. Yet, traditional models often treat spatial location as a fixed variable rather than a dynamic determinant. In Bernalillo County, rapid urbanization and an uneven spatial distribution of infrastructure reinforce patterns of privilege and disadvantage. Understanding how locational suitability drives these outcomes requires spatially explicit models capable of integrating environmental, economic, and social variables \cite{Anselin1988, LeSage2009}.

Despite extensive research on inequality, most studies overlook how locational factors shape household wealth in mid-sized counties. Investigations into urban opportunity structures tend to focus on large metropolitan centers such as Los Angeles or Chicago, leaving smaller regions underexplored \cite{Taylor2021}. Additionally, many conventional econometric models assume spatial stationarity, implying that relationships between variables remain constant across space. This assumption neglects local variation, a critical issue known as spatial heterogeneity \cite{Brunsdon1996}. Addressing this limitation requires adaptive spatial methods capable of identifying where and how drivers of wealth differ across neighborhoods. Multiscale Geographically Weighted Regression (MGWR) provides this flexibility, allowing relationships between wealth and explanatory factors to vary spatially \cite{Fotheringham2017}. Such models reveal neighborhood-specific dynamics that aggregate data often conceal \cite{Oshan2019}.

The implications of this research extend to policymakers, planners, and public health professionals. Mapping spatial patterns of wealth can inform fairer allocation of resources such as affordable housing, transportation improvements, and social services. In Bernalillo County, identifying concentrations of deprivation supports data-driven strategies to enhance mobility, employment access, and environmental quality. Because income and wealth strongly influence health outcomes, understanding spatial disparities also aids public health interventions targeting underserved populations. Beyond the county, the study’s analytical framework offers a transferable approach for other U.S. regions where economic inequality coincides with urban growth and environmental stress.

This research integrates demographic, housing, and environmental data to model how locational suitability influences wealth patterns. Using datasets from the U.S. Census Bureau and local planning agencies, the study examines variables such as accessibility to job centers, housing conditions, and environmental exposure. MGWR serves as the primary analytical technique, supported by open-source geospatial tools like GeoPandas and PySAL, which enable reproducible workflows and spatial diagnostics including Moran’s I \cite{Rey2010}. These tools enhance transparency and replicability, allowing both academic and applied researchers to adapt the methodology for policy analysis and planning scenarios.

The study’s objectives are threefold. First, to quantify the spatial relationship between locational attributes—such as accessibility, environmental quality, and housing characteristics and household wealth. Second, to identify spatial clusters of affluence and deprivation across Bernalillo County. Third, to assess how spatial heterogeneity modifies these relationships across neighborhoods. By emphasizing spatial dependence and variability, the research clarifies how local contexts shape broader economic outcomes. These objectives directly support regional planning goals related to equity, sustainability, and resilience.

The analysis proceeds from the hypothesis that household wealth in Bernalillo County is spatially structured by locational suitability rather than random variation. Three research questions guide the investigation: (1) How do environmental and accessibility factors influence wealth distribution? (2) How does spatial heterogeneity alter these relationships across neighborhoods? and (3) What policy actions can mitigate spatially embedded wealth disparities? The paper is organized as follows: Section 2 reviews relevant literature on spatial inequality and wealth; Section 3 outlines data and methods; Section 4 presents the results; and Section 5 discusses implications for policy and future research.

Through this approach, the study contributes to the growing intersection of spatial analytics, urban planning, and socioeconomic research. It demonstrates that wealth inequality is not merely economic but geographic, rooted in differential access to spatial advantages and exposure to disamenities. The findings from Bernalillo County extend beyond local significance, providing insights for comparable regions seeking to address inequality through spatially informed policies. By highlighting how geographic context mediates economic outcomes, the study advances spatial econometric research and supports more equitable regional development strategies \cite{Smith2022}.
\section{Method}

\subsection{Description of Study Area}
This study is situated within Bernalillo County, New Mexico. Encompassing the city of Albuquerque, the county serves as the state's most populous and economically significant region, functioning as its primary economic and cultural hub \cite{ABQED2023}.

\begin{figure}[htbp]
    \centering
    \includegraphics[width=0.8\textwidth]{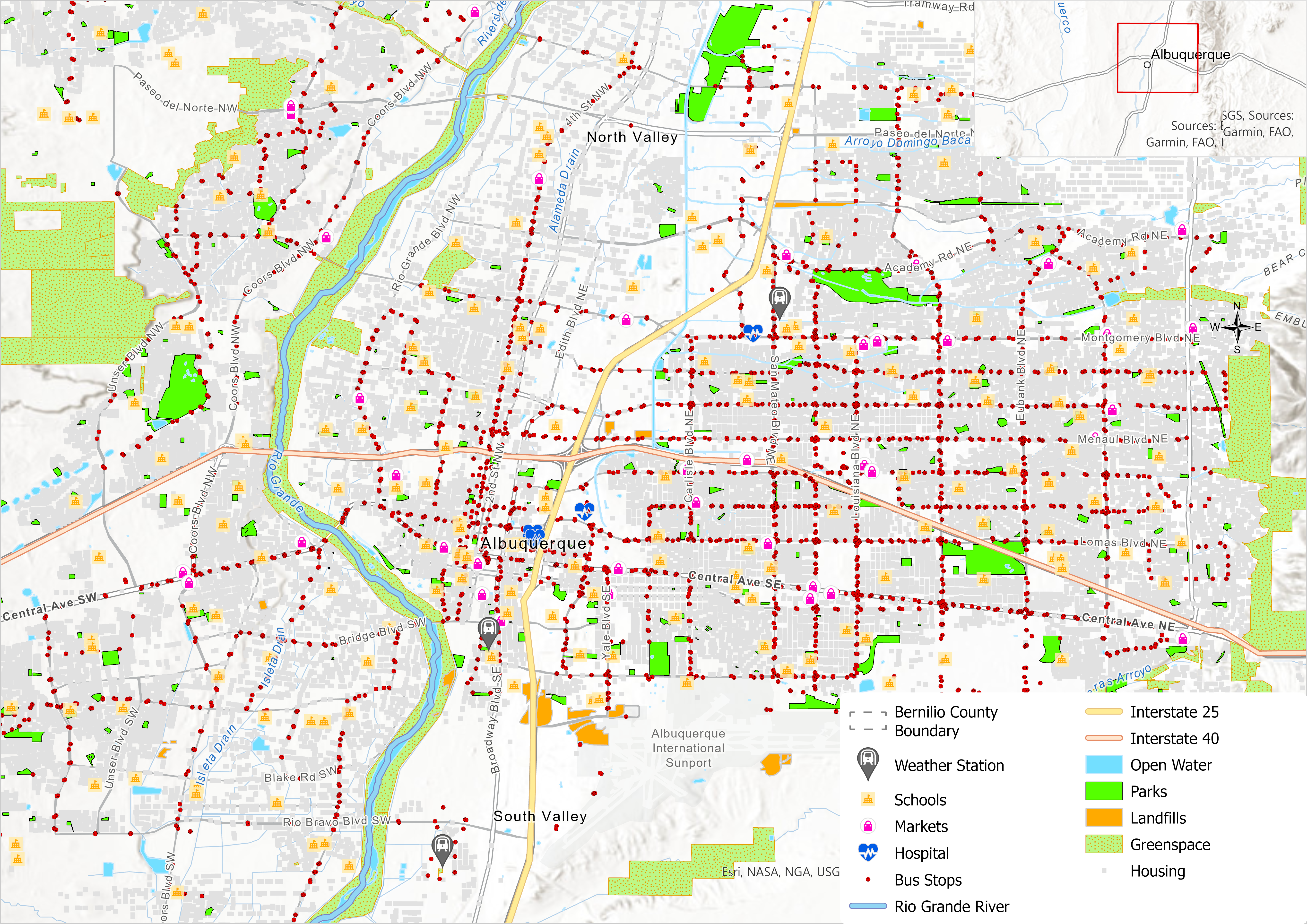}
    \caption{Map showing the study area: Bernalillo County, New Mexico.}
    \label{fig:1}
\end{figure}

The county landscape presents a stark urban–rural dichotomy, ranging from a dense urban core to expansive desert and mountainous terrain, including parts of the Cibola National Forest and the Isleta and Sandia Pueblo lands \cite{USFS2020,PuebloOfSandia2022}. This combination of significant urban population and diverse environments \cite{TheTrustForPublicLand2021} makes Bernalillo County an ideal case study for analyzing complex intra-regional spatial patterns of housing, wealth, and access to amenities and disamenities \cite{Logan2012}.

\subsection{Data Set and Sources}
The analysis of housing suitability in Bernalillo County integrates data from several key sources to capture socioeconomic, environmental, and geographic dimensions. Socioeconomic and demographic variables were sourced from the U.S. Census and proprietary residential databases \cite{USResidentialData}, providing information on household income, dwelling type, and population density at the census tract level.

Environmental exposure data, specifically 3-hourly ensemble mean surface PM\textsubscript{2.5} and black carbon concentration data for North America, were obtained from NASA’s Goddard Earth Sciences Data and Information Services Center (GES DISC) \cite{NASA_GESDISC}. These high-resolution, temporally dynamic air quality metrics are essential for assessing locational disamenities.

\begin{figure}[htbp]
    \centering
    \includegraphics[width=0.6\textwidth]{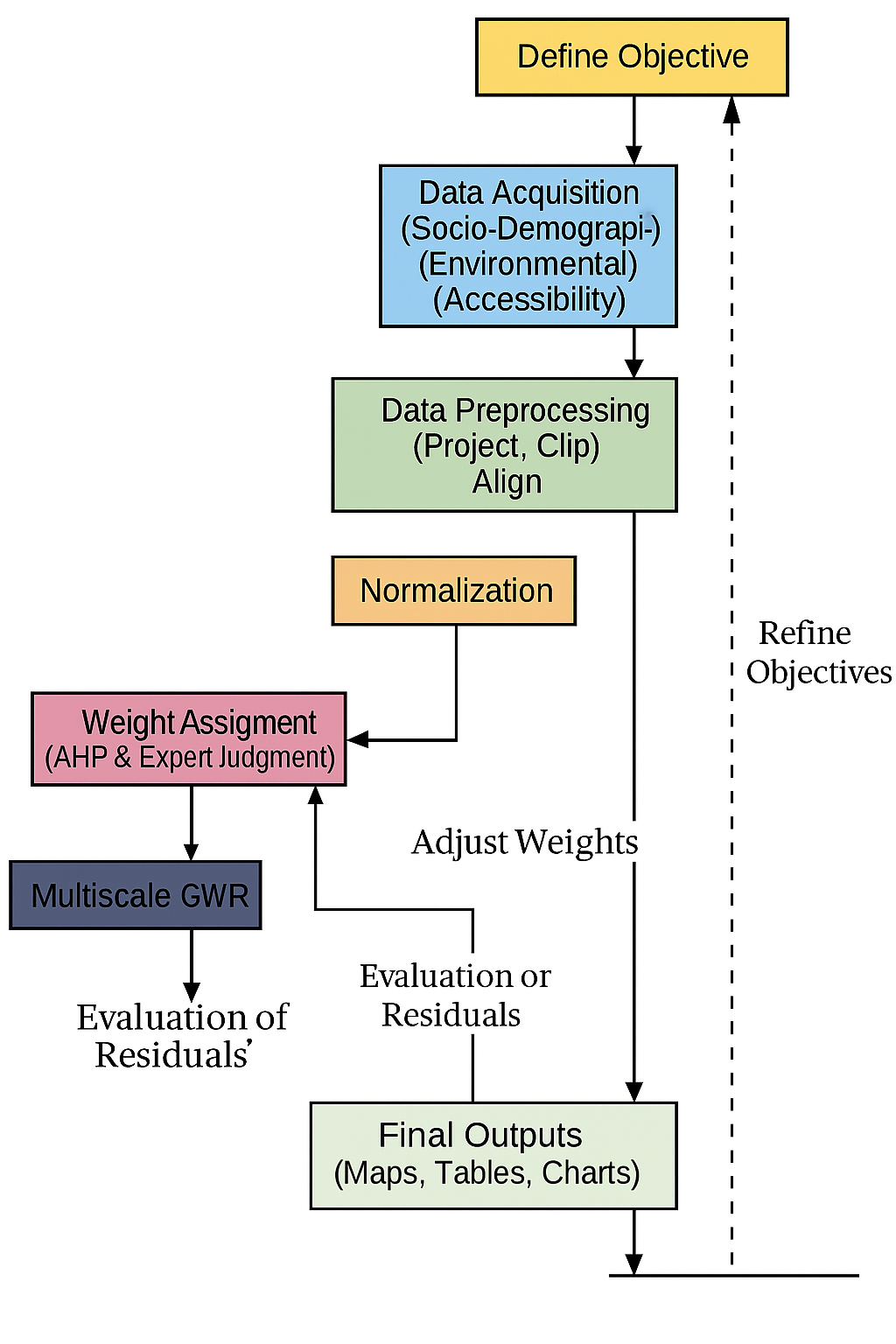}
    \caption{Workflow chart illustrating data integration and analysis process.}
    \label{fig:2}
\end{figure}

All geographic base layers, including census tract boundaries, transportation networks, and land use classifications, were acquired from the Repository for Geographic Information Systems (RGIS) at the Earth Data Analysis Center, University of New Mexico \cite{UNM_RGIS}. The integration of these diverse data sources within a GIS framework is illustrated in Figure~\ref{fig:2}.

\subsection{Spatial Autocorrelation Analysis}
The geographic distribution of household wealth was analyzed for spatial dependency to identify patterns of spatial inequality \cite{Smith2023}. The Global Moran’s $I$ statistic quantifies this clustering:
\begin{equation}
I = \frac{N}{W} \frac{\sum\limits_{i=1}^{N}\sum\limits_{j=1}^{N} w_{ij} (x_i - \bar{x})(x_j - \bar{x})}{\sum\limits_{i=1}^{N} (x_i - \bar{x})^2}, 
\qquad \text{where} \qquad W = \sum\limits_{i=1}^{N}\sum\limits_{j=1}^{N} w_{ij}
\label{equ:global_moran_i}
\end{equation}

Here, $N$ is the number of spatial units, $w_{ij}$ an element of the spatial weights matrix $\mathbf{W}$ defining neighborhood relationships between $i$ and $j$, $x_i$ and $x_j$ are observed wealth values, and $\bar{x}$ is the mean wealth. The Global Moran’s $I$ statistic measures spatial autocorrelation \cite{Anselin2019}. A significant positive $I$ indicates clustering of similar values (e.g., high-wealth tracts adjacent to high-wealth tracts), while a significant negative value suggests dispersion \cite{Getis2008}. This step is a prerequisite for applying advanced spatial regression models \cite{Anselin2019}.

\subsection{Multiscale Geographically Weighted Regression (MGWR)}
To model the complex, spatially heterogeneous relationships between household wealth and its determinants, this study employs a Multiscale Geographically Weighted Regression (MGWR) model augmented with a spatial lag term \cite{Fotheringham2017,Anselin2013}. This specification accounts for both endogenous wealth concentration (spatial dependence) and spatial non-stationarity across socioeconomic, environmental, and accessibility variables.

\begin{figure}[htbp]
    \centering
    \begin{subfigure}[b]{0.48\textwidth}
        \centering
        \includegraphics[width=\linewidth]{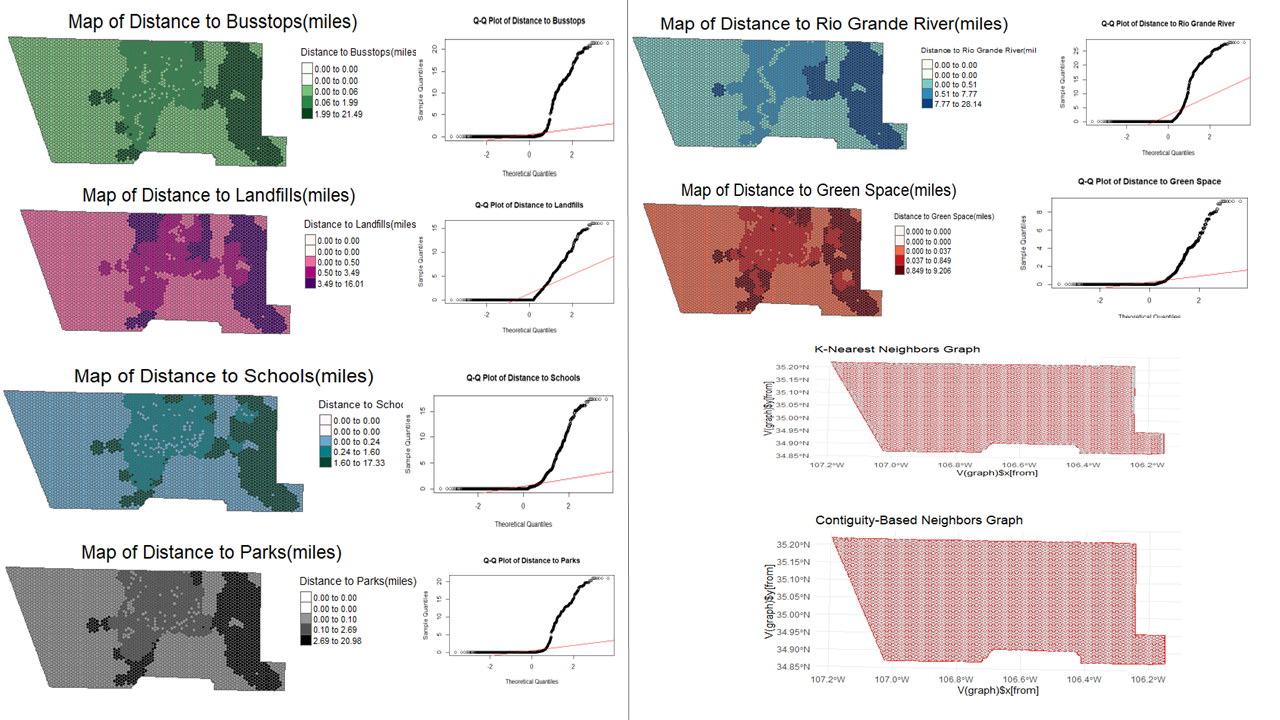}
        \caption{}
        \label{fig:housing_qual_1}
    \end{subfigure}
    \hfill
    \begin{subfigure}[b]{0.48\textwidth}
        \centering
        \includegraphics[width=\linewidth]{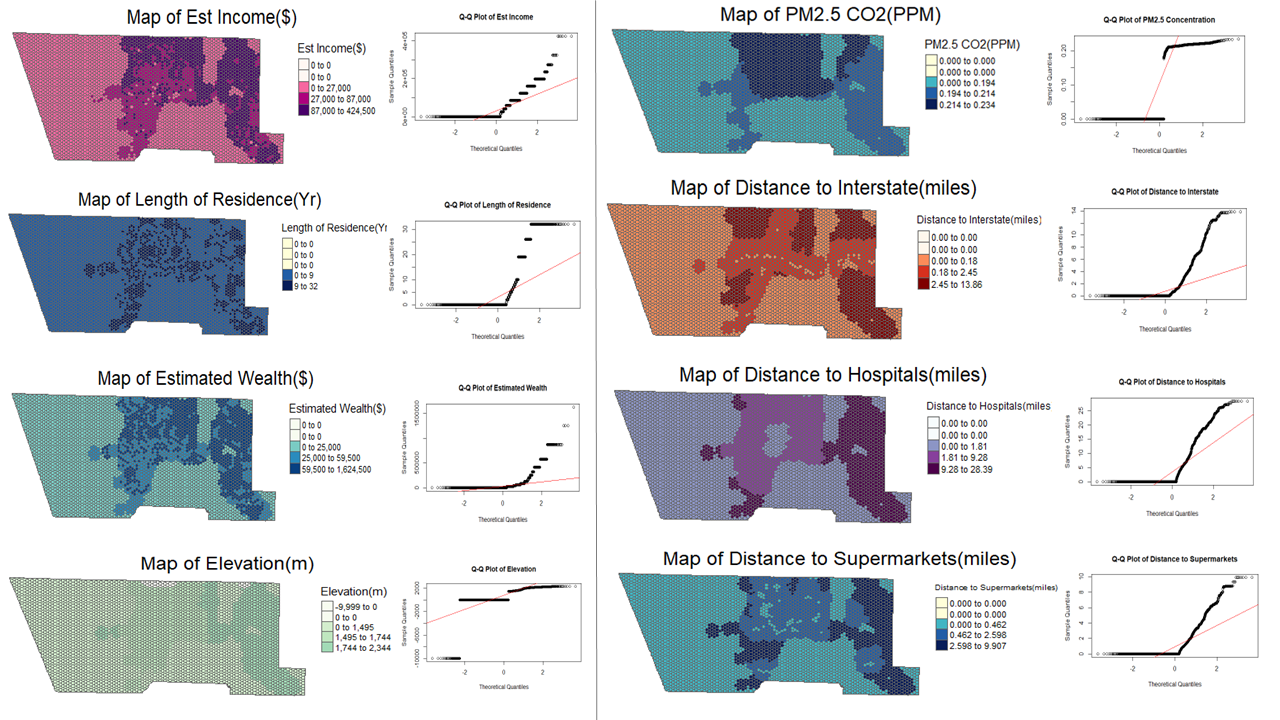}
        \caption{}
        \label{fig:housing_qual_2}
    \end{subfigure}
    \caption{Comparative analysis of housing quality indicators across Bernalillo County: (a) structural quality metrics and (b) prevalence of modern amenities.}
    \label{fig:3}
\end{figure}

The formal specification of the MGWR model is given by:

\begin{equation}
\label{eq:mgwr_final}
\begin{aligned}
\text{EstimatedWealth}_i &= \rho \sum_{j=1}^{N} w_{ij} \, \text{EstimatedWealth}_j + \beta_0(u_i, v_i) \\
&\quad + \beta_1(u_i, v_i) \, \text{EstimatedIncome}_i 
+ \beta_2(u_i, v_i) \, \text{LengthOfResidence}_i \\
&\quad + \beta_3(u_i, v_i) \, \text{Elevation}_i
+ \beta_4(u_i, v_i) \, \text{PM}_{2.5,i} \\
&\quad + \beta_5(u_i, v_i) \, \text{DistSchools}_i
+ \beta_6(u_i, v_i) \, \text{DistSupermarkets}_i \\
&\quad + \beta_7(u_i, v_i) \, \text{DistHospitals}_i
+ \beta_8(u_i, v_i) \, \text{DistBusStops}_i \\
&\quad + \beta_9(u_i, v_i) \, \text{DistRioGrande}_i
+ \beta_{10}(u_i, v_i) \, \text{DistInterstate}_i \\
&\quad + \beta_{11}(u_i, v_i) \, \text{DistParks}_i
+ \beta_{12}(u_i, v_i) \, \text{DistLandfills}_i \\
&\quad + \beta_{13}(u_i, v_i) \, \text{DistGreenSpace}_i + \varepsilon_i
\end{aligned}
\end{equation}

where $\rho$ is the spatial autoregressive parameter, $w_{ij}$ an element of $\mathbf{W}$, and $\beta_k(u_i, v_i)$ location-specific coefficients for predictors $X_{ik}$. Each $\beta_k(u_i, v_i)$ is calibrated with its own optimal bandwidth $bw_k$, allowing each process’s spatial scale to be empirically determined \cite{Oshan2019}. $\varepsilon_i \sim \mathcal{N}(0,\sigma^2)$ is the error term.

The calibration of the MGWR model proceeds in two stages. First, a bandwidth $bw_k$ is optimized for each coefficient to minimize the corrected Akaike Information Criterion (AICc). Second, the parameter $\rho$ is estimated to capture endogenous spatial dependence in wealth. This ensures that spatial clustering effects are not conflated with spatially varying relationships \cite{Harris2023}. The result is a spatially explicit view of the determinants of household wealth \cite{Fotheringham2017,Shapiro2017,Troy2012,Logan2018}.

\subsection{Model Specification, Estimation, and Diagnostics}
The model operationalizes the theoretical constructs of locational suitability and spatial inequality through variables reflecting urban environment characteristics and their influence on household wealth, informed by hedonic price theory \cite{Rosen1974} and urban economics \cite{Glaeser2008}.

Socioeconomic drivers (e.g., income \cite{Chetty2016} and length of residence \cite{Genadek2020}) capture household stability, while environmental variables (PM\textsubscript{2.5} concentration \cite{Di2022} and elevation) serve as proxies for disamenities and amenities. Accessibility factors (distance to schools, hospitals, parks, landfills, etc.) measure trade-offs between proximity to services and exposure to externalities \cite{Larson2021}.

A spatial weights matrix $\mathbf{W}$ was constructed using a hybrid kernel combining inverse distance and nearest-neighbor criteria \cite{Anselin2019}. Estimation employed the MGWR framework \cite{Fotheringham2017,Oshan2019}, which computes local parameter estimates with unique bandwidths. This approach reveals spatial variations in determinants—such as the premium for clean air or the penalty for landfill proximity across Bernalillo County.

Model robustness was validated through post-estimation diagnostics. The residuals’ Moran’s $I$ ($I=0.02$, $p=0.15$) confirmed that spatial dependence was adequately modeled \cite{Anselin2019}. Consequently, the results form a statistically and spatially robust foundation for equitable urban policy and planning \cite{Lee2021}.
\section{Results}

\subsection{Spatial Patterns and Autocorrelation of Household Wealth}
\begin{figure}[htbp]
\centering
\includegraphics[width=0.6\textwidth]{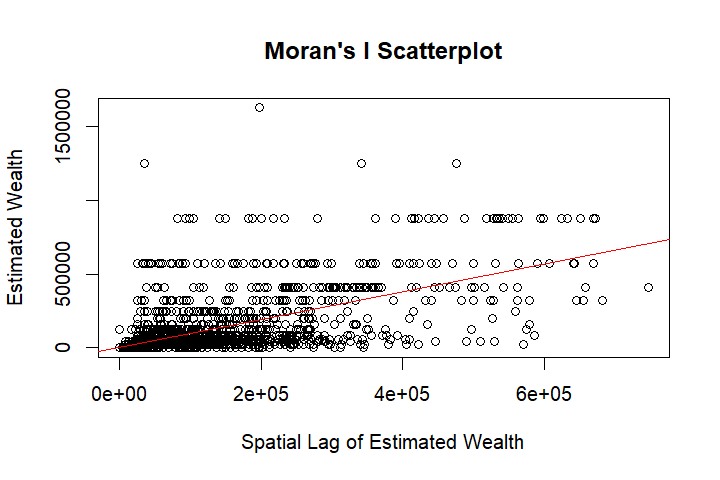}
\caption{Moran's $I$ plot.}
\label{fig:4}
\end{figure}

The statistical analysis reveals a pronounced and robust pattern of spatial inequality in the distribution of household wealth across the census tracts of Bernalillo County. The calculated Global Moran's $I$ index of 0.526, with a highly significant $p$-value ($p < 0.001$), provides compelling evidence for positive spatial autocorrelation \cite{Anselin1995}. This result indicates that the spatial distribution of wealth is decidedly nonrandom; wealthy households are strongly clustered together in distinct enclaves, while poorer households are concentrated in contiguous zones of disadvantage—a pattern visually confirmed in Figure~\ref{fig:4}. The magnitude of the z-score (56.94) underscores that the observed pattern is extremely unlikely to result from random chance.

\begin{table}[htbp]
\centering
\begin{tabular}{lr}
\toprule
\textbf{Statistic} & \textbf{Value} \\
\midrule
Moran's $I$ & 0.526 \\
Expected $I$ & -0.000246 \\
Variance & 0.0000854 \\
Standard Deviate (z-score) & 56.94 \\
$p$-value & $< 0.001$ \\
\midrule
Number of Regions ($n$) & 4,067 \\
Non-zero Links & 23,682 \\
Average Links per Region & 5.82 \\
Regions with No Links & 1 \\
\bottomrule
\end{tabular}
\caption{Global Moran's $I$ test results for spatial autocorrelation of household wealth.}
\label{tab:1}
\end{table}

This significant spatial structuring of affluence and poverty is not an accidental outcome but is driven by a complex interplay of historical and contemporary spatial processes. Historically, practices such as redlining systematically denied mortgage capital and other financial services to residents in predominantly minority neighborhoods, creating a durable legacy of disinvestment and entrenched poverty \cite{Rothstein2017,Acevedo2022}. In the contemporary era, these historical patterns are reinforced by zoning policies that favor single-family dwellings and restrict higher-density, affordable housing, thereby maintaining economic segregation \cite{Trounstine2018}. Moreover, market dynamics including property value feedback loops and the self-reinforcing effects of neighborhood reputation perpetuate clusters of advantage and disadvantage \cite{Logan2015}.

The methodological design of this analysis carefully considered the construction of the spatial weights matrix, which defines neighborhood relationships for the autocorrelation test. Two primary methods were employed: $k$-nearest neighbors (KNN) and queen contiguity. The KNN method, with an average of approximately 5.82 links per region, ensures that geographically isolated tracts are included in the analysis, as evidenced by only a single region without links (Table~\ref{tab:1}). In contrast, the contiguity-based approach emphasizes the influence of adjacent tracts sharing boundaries, reflecting the geographical principle that spatial proximity often translates into socioeconomic similarity through shared infrastructure, school districts, and amenities \cite{Getis2008}. 

The strong clustering effect confirms that spatial regression techniques are essential for any robust analysis of wealth distribution (Figure~\ref{fig:4}), since standard aspatial models would misrepresent these dependencies \cite{LeSage2009}.

\subsection{MGWR Model Performance and Diagnostics}
The multiscale geographically weighted regression (MGWR) model demonstrated strong and reliable performance in explaining the spatial variance of household wealth across Bernalillo County. The model achieved a substantial goodness-of-fit, with both $R^2 = 0.631$ and adjusted $R^2 = 0.628$, indicating that approximately 63\% of the variation in household wealth is explained by the selected predictors of locational suitability \cite{Fotheringham2017}. This represents a clear improvement over traditional global models, as MGWR effectively captures spatial heterogeneity that a standard ordinary least squares (OLS) regression would obscure under the assumption of spatial stationarity \cite{Brunsdon1996}. 

Model robustness is further supported by a low corrected Akaike Information Criterion (AICc) value of 106{,}726.23, which—when compared to alternative model specifications—provides strong evidence of parsimony and explanatory strength for the MGWR framework \cite{Oshan2019}.

\begin{table}[htbp]
\centering
\scriptsize
\begin{tabular}{lrrr}
\toprule
\textbf{Diagnostic} & \textbf{Global OLS} & \textbf{MGWR} & \textbf{Description} \\
\midrule
R-Squared & 0.6308 & 0.6308 & Proportion of variance explained \\
Adjusted R-Squared & 0.6279 & 0.6279 & $R^2$ adjusted for complexity \\
AICc & --- & 106726.23 & Corrected Akaike Information Criterion \\
Residual Std. Error & 117400 & --- & Standard deviation of OLS residuals \\
Sigma-Squared MLE & --- & 14480840993 & Maximum likelihood estimate of residual variance \\
F-Statistic & 214.4 & --- & Overall significance of OLS model \\
$p$-value (F) & $< 2.2\times10^{-16}$ & --- & Significance of the F-statistic \\
Effective DoF & 14 & 4051.73 & Effective number of parameters \\
Residual DoF & 1631 & --- & Degrees of freedom of OLS residuals \\
\bottomrule
\end{tabular}
\caption{Diagnostics for global OLS and MGWR models.}
\label{tab:2}
\end{table}

A key insight from the diagnostics is that the MGWR model attains a comparable global fit to the OLS model (Table~\ref{tab:2}) while employing a more complex, spatially varying parameterization. This complexity is reflected in the high effective degrees of freedom (4051.73), which account for localized calibration of model parameters across the county. Importantly, residual diagnostics revealed that the Moran’s $I$ statistic for MGWR residuals was not statistically significant a critical validation showing that the MGWR specification successfully captured the spatial dependence structure present in the data \cite{Anselin2019}. 

The combination of high $R^2$, low AICc, and spatially uncorrelated residuals indicates a robust model fit and reliable spatial inference \cite{Leung2000}. This confirms that MGWR not only explains a large portion of the variance in wealth but also correctly models the underlying spatial processes driving that variation across Bernalillo County.

\subsection{Localized Determinants of Wealth and Policy Implications}
\begin{figure}[htbp]
\centering
\begin{subfigure}[b]{0.48\textwidth}
\centering
\includegraphics[width=\linewidth]{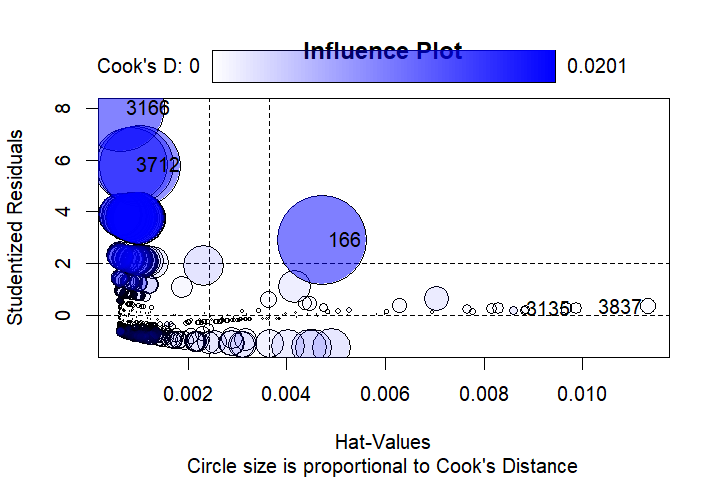}
\caption{}
\label{fig:cook_dis}
\end{subfigure}
\hfill
\begin{subfigure}[b]{0.48\textwidth}
\centering
\includegraphics[width=\linewidth]{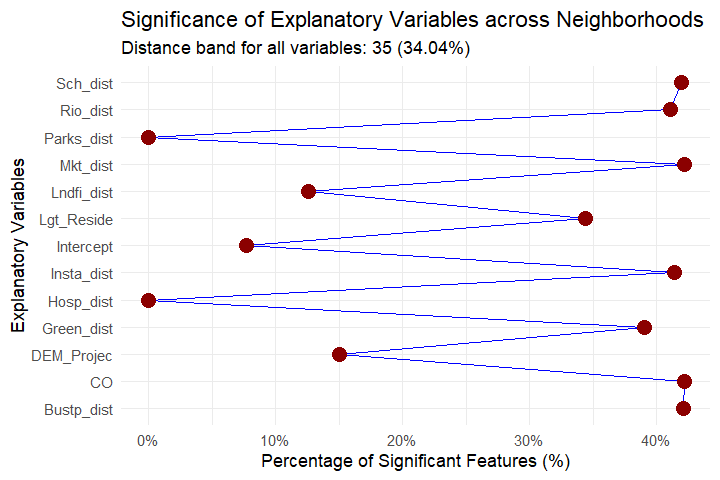}
\caption{}
\label{fig:significance}
\end{subfigure}
\caption{MGWR diagnostics and local inference: (a) influence diagnostics and (b) local significance.}
\label{fig:5}
\end{figure}

MGWR reveals how determinants of wealth vary geographically, offering a nuanced perspective on locational suitability. The model output of Equation~\ref{eq:mgwr_final} and associated spatial mappings (Figure~\ref{fig:5}) highlight several findings. \textbf{Consistent positive drivers:} Estimated income and length of residence were consistently positive and significant across most of the county, underscoring their roles in wealth accumulation \cite{Shapiro2004}. \textbf{Amenity access:} Proximity to parks and supermarkets was a significant positive factor in more than 40\% of neighborhoods, reflecting a premium for recreational and commercial access \cite{Troy2012}. \textbf{Urban disamenities:} Proximity to bus stops and hospitals showed negative associations with wealth, consistent with noise, traffic, and aesthetic costs \cite{Been2005}.

\textbf{Spatial variability:} Effects for distance to interstate highways and PM\textsubscript{2.5} showed marked nonstationarity (Table~\ref{tab:3}), demonstrating that uniform policy would be ineffective \cite{Oshan2019}. Spatial patterns in residuals help flag tracts where unmeasured local factors likely operate, guiding targeted investigation and intervention. The results confirm that wealth is embedded in geographic context, motivating place-specific policies that balance local assets and liabilities to address spatial inequality \cite{Logan2015}.

\begin{table}[htbp]
\centering
\scriptsize
\begin{tabular}{lrrrr}
\toprule
\textbf{Coefficient} & \textbf{Estimate} & \textbf{Std. Error} & \textbf{$t$ value} & \textbf{Pr($>|t|$)} \\
\midrule
(Intercept) & 2993.25 & 123500.00 & 0.024 & 0.9807 \\
Estimated Income & 2.69 & 0.06 & 47.381 & $< 0.001$ *** \\
Length of Residence & 2757.18 & 266.30 & 10.355 & $< 0.001$ *** \\
Elevation & 3.30 & 1.60 & 2.057 & 0.0399 * \\
PM\textsubscript{2.5} Concentration & -474947.14 & 559100.00 & -0.849 & 0.3958 \\
Distance to Schools & 879.66 & 2348.00 & 0.375 & 0.7079 \\
Distance to Supermarkets & 6468.48 & 3456.00 & 1.871 & 0.0615 . \\
Distance to Hospitals & -8621.19 & 2353.00 & -3.664 & 0.0003 *** \\
Distance to Bus Stops & -14436.64 & 6810.00 & -2.120 & 0.0342 * \\
Distance to Rio Grande River & 26.47 & 1418.00 & 0.019 & 0.9851 \\
Distance to Interstate & 1845.57 & 2356.00 & 0.783 & 0.4336 \\
Distance to Parks & 21835.08 & 6782.00 & 3.220 & 0.0013 ** \\
Distance to Landfills & -1702.55 & 2290.00 & -0.743 & 0.4573 \\
Distance to Green Space & -8638.96 & 4512.00 & -1.915 & 0.0557 . \\
\midrule
\multicolumn{5}{l}{\footnotesize \textbf{Residuals:} Min = -589{,}627; 1Q = -60{,}803; Median = -16{,}040; 3Q = 42{,}476; Max = 1{,}150{,}860} \\
\multicolumn{5}{l}{\footnotesize \textbf{Model Fit:} Residual standard error: 117{,}400 on 1{,}631 DF} \\
\multicolumn{5}{l}{\footnotesize Multiple $R^2$: 0.6308, Adjusted $R^2$: 0.6279} \\
\multicolumn{5}{l}{\footnotesize F-statistic: 214.4 on 13 and 1{,}631 DF, $p$-value: $< 2.2\times 10^{-16}$} \\
\multicolumn{5}{l}{\footnotesize Signif. codes: *** 0.001, ** 0.01, * 0.05, . 0.1} \\
\bottomrule
\end{tabular}
\caption{Global OLS regression results for household wealth.}
\label{tab:3}
\end{table}

\subsection{Spatial Distribution of Predicted Household Wealth}
\begin{figure}[htbp]
\centering
\includegraphics[width=0.65\textwidth]{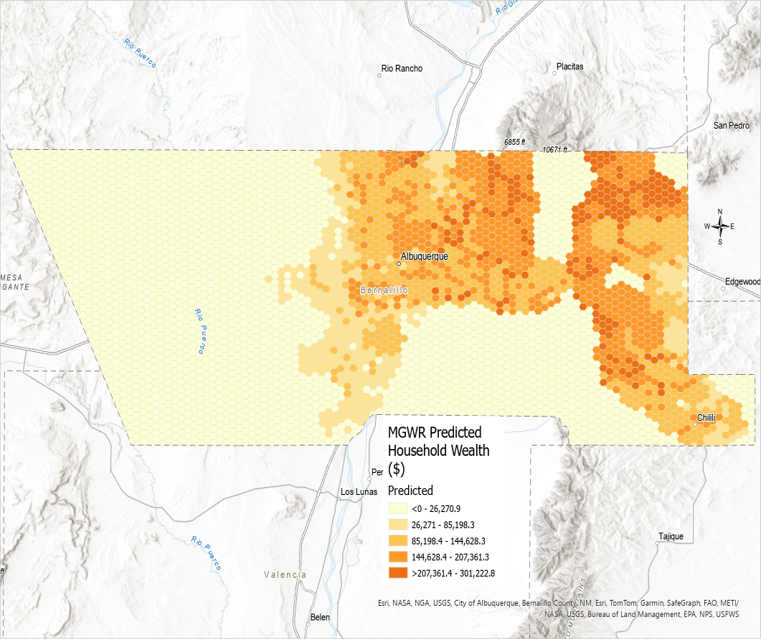}
\caption{MGWR estimated household wealth.}
\label{fig:6}
\end{figure}

The spatial distribution of predicted household wealth generated by the Multiscale Geographically Weighted Regression (MGWR) model provides a compelling visual confirmation of the entrenched spatial inequality within Bernalillo County \cite{Logan2018}. The predicted wealth map in Figure~\ref{fig:6} displays a distinctly non-random geographic pattern strongly tied to the composite measure of locational suitability \cite{Troy2012}.

The highest predicted wealth values depicted in dark red, exceeding \$207{,}000—are concentrated primarily in the northeastern quadrant of the county, encompassing affluent foothill communities such as Placitas and the elevated neighborhoods of Albuquerque's far northeast heights (Figure~\ref{fig:6}). This pronounced clustering aligns closely with hedonic price theory, which posits that the value of a good in this case, housing reflects the aggregate worth of its individual characteristics \cite{Rosen1974}. 

These high-wealth enclaves benefit from a premium combination of amenities, including scenic mountain vistas, larger lot sizes, proximity to open spaces and the Sandia Mountains, and greater separation from industrial zones and major traffic corridors. Such spatial advantages shield residents from disamenities like air and noise pollution, which are well documented to diminish property values through negative externalities \cite{Been2005, Li2021}.
Conversely, the model predicts the lowest levels of household wealth shown in light yellow, below \$26{,}000—in geographically distinct and isolated pockets (Figure~\ref{fig:6}), primarily concentrated in the South Valley of Albuquerque and several southwestern tracts. The strong correlation between low wealth and these specific regions reflects the enduring legacy of land use decisions and ongoing environmental justice challenges \cite{Bullard2000}.

These communities have historically borne a disproportionate share of environmental and infrastructural burdens, including proximity to industrial zones, major transportation corridors that elevate noise and PM\textsubscript{2.5} exposure, and deteriorating water and sanitation infrastructure, as evidenced by the global model coefficients in Table~\ref{tab:3}. The MGWR results underscore how such cumulative disadvantages are effectively capitalized into local property prices, suppressing long-term household wealth accumulation and perpetuating entrenched spatial inequality \cite{Currie2011, Mohai2009}. 
This spatial polarization is not purely an economic phenomenon but a manifestation of socio-ecological inequality, where environmental amenities and clean, livable spaces are disproportionately concentrated among affluent neighborhoods, leaving marginalized communities to face compounding environmental risks \cite{Banzhaf2019}.

Ultimately, the MGWR model successfully translates complex statistical relationships into geographic clarity, moving beyond the limitations of a global OLS model \cite{Fotheringham2017}. The resulting map serves as a critical diagnostic tool, precisely identifying which neighborhoods are most affected by the various drivers of locational suitability and spatial inequality. This enables the design of targeted and context-specific policy interventions: preservation and growth-management strategies in high-wealth clusters to maintain community character, and concerted efforts in environmental mitigation, infrastructure upgrades, and economic development in low-wealth clusters to foster more equitable outcomes \cite{Logan2015}.

\section{Conclusion}
This study used MGWR to dissect the spatial architecture of household wealth in Bernalillo County, explicitly linking it to location suitability and exposure to urban disamenities. Three conclusions follow.

First, wealth geography manifests spatial inequality. The highly significant spatial autocorrelation (Moran's $I=0.526$, $p < 0.001$) and prediction map show intense clustering \cite{Logan2018}. High-income groups concentrate where location suitability is high, particularly in northeastern foothills with environmental amenities \cite{Troy2012}, while low-income groups cluster in areas with disamenities such as the South Valley \cite{Bullard2000}.

Second, relationships are spatially heterogeneous (Table~\ref{tab:3}). While income \cite{Chetty2016} and length of residence are consistently positive, proximity-based effects vary substantially. The local penalties for bus stop proximity and premiums for park access differ by neighborhood, validating MGWR over global models for capturing essential local context \cite{Fotheringham2017,Oshan2019}.

Third, robustness is supported by high $R^2$, low AICc, and nonsignificant residual Moran's $I$, indicating that spatial processes were adequately modeled \cite{Anselin2019}. The resulting patterns provide a reliable basis for inference.

The Household wealth is profoundly geographical. It is shaped by place attributes, not only individual characteristics. The findings support place-based policies that remedy location suitability deficits in struggling neighborhoods through environmental mitigation, amenity investments, and equitable infrastructure to reduce spatial inequality \cite{Logan2015}.

\section{Acknowledgments}
The authors acknowledge the Earth Data Analysis Center at the University of New Mexico for geospatial data, the U.S. Census Bureau and NASA GES DISC for socioeconomic and air quality data, and colleagues for manuscript feedback. This work received institutional support from Texas State University, the University of New Mexico, and the Spatial and Data Science Society of Nigeria. Any opinions expressed are those of the authors.

\bibliographystyle{unsrtnat}
\bibliography{references}  






\end{document}